%% file: main.tex
\documentclass[journal]{IEEEtran}
\usepackage[left=1.43cm,right=1.43cm,top=1.8cm,bottom=4.21cm]{geometry}
\usepackage[algo2e]{algorithm2e} 
\usepackage{amsmath,amssymb,amsmath,amsfonts}
\usepackage{bm}
\usepackage{bbm}
\usepackage{dsfont}
\usepackage{graphicx}
\usepackage{cite}
\usepackage{epstopdf}
\usepackage{gensymb}
\usepackage{color}
\usepackage{lipsum}
\usepackage{float}
\usepackage{epstopdf}
\usepackage[mathcal]{eucal}
\usepackage{subfiles}
\usepackage[T1]{fontenc}
\usepackage{siunitx}
\usepackage{tabulary}
\usepackage{epsfig,graphics,graphicx,subfig,amssymb,amstext,amsmath,algorithm,algorithmic,wrapfig,multirow}
\usepackage{array}
\usepackage{adjustbox}
\usepackage[dvipsnames]{xcolor}
\usepackage{cite}
\usepackage{times}
\usepackage{placeins}
\usepackage{textcomp}
\usepackage[font=small]{caption}
\usepackage{flushend}
\usepackage{color, colortbl}
\usepackage{tabularx}
\usepackage{bm}
\usepackage{enumerate}
\usepackage{tikz}
\usepackage{pgfplots}
\usepackage{epstopdf}
\usetikzlibrary{shapes,arrows}
\usepackage[utf8]{inputenc}
\usepackage{mathtools}
\usepackage[utf8]{inputenc}
\usepackage{xcolor}
\usepackage{tikz}
\usetikzlibrary{chains,arrows,calc,positioning}
\usepackage{amssymb}
\usepackage{pifont}
\usepackage{soul}
\usepackage{breqn} 
\usepackage{booktabs} 
\usepackage{multirow}
\usepackage{enumitem}
\usepackage{tabulary}
\usepackage[normalem]{ulem}
\usepackage{dblfloatfix}
\usepackage{makecell}
\usepackage{lipsum}
\usepackage{amsthm}
\usepackage{comment}
\usepackage{filecontents}
\usepackage{bbm}

\newtheoremstyle{mystyle}
  {}
  {}
  {\itshape}
  {}
  {\bfseries}
  {.}
  { }
  {}
\theoremstyle{mystyle}

\newlength \figwidth
\definecolor{bittersweet}{rgb}{1.0, 0.44, 0.37}
\definecolor{glaucous}{rgb}{0.38, 0.51, 0.71}
\definecolor{gainsboro}{rgb}{0.86, 0.86, 0.86}
\definecolor{babyblueeyes}{rgb}{0.63, 0.79, 0.95}
\definecolor{silver}{rgb}{0.75, 0.75, 0.75}
\definecolor{neoncarrot}{rgb}{1.0, 0.64, 0.26}
\definecolor{Gray}{gray}{0.9}
\definecolor{LightCyan}{rgb}{0.88,1,1}
\definecolor{BackgroundLightBlue}{rgb}{0.97,0.97,1}
\definecolor{BackgroundGray}{gray}{0.98}

\makeatletter

 \let\oldforeign@language\foreign@language
 \DeclareRobustCommand{\foreign@language}[1]{%
   \lowercase{\oldforeign@language{#1}}}

\input{commands.tex}

\makeatother

\IEEEoverridecommandlockouts
\begin{document}

\bstctlcite{IEEEexample:BSTcontrol}

\title{Data-Driven Cellular Mobility Management via Bayesian Optimization and Reinforcement Learning}

\author{\IEEEauthorblockN{{Mohamed Benzaghta, Sahar Ammar, David L\'{o}pez-P\'{e}rez}, Basem Shihada, and Giovanni Geraci} 
\thanks{M.~Benzaghta is with Universitat Pompeu Fabra, Spain.}
\thanks{S.~Ammar and B.~Shihada are with King Abdullah University of Science and Technology, Saudi Arabia.}
\thanks{D.~López-Pérez is with  Universitat Politècnica de Val{è}ncia, Spain.}
\thanks{G.~Geraci is with Nokia Standards and Universitat Pompeu Fabra, Spain. He was with Telefónica Research, Spain, when the work was carried out.
} 
\thanks{This work was supported by 
\emph{a)} HORIZON-SESAR-2023-DES-ER-02 project ANTENNAE (101167288),
\emph{b)} the Spanish Ministry of Economic Affairs and Digital Transformation and the European Union NextGenerationEU through actions CNS2023-145384 and CNS2023-144333, 
\emph{c)} the Spanish State Research Agency through grants PID2021-123999OB-I00 and CEX2021-001195-M, 
\emph{d)} the Generalitat Valenciana, Spain, through the  CIDEGENT PlaGenT, Grant CIDEXG/2022/17, Project iTENTE, and \emph{e)} the UPF-Fractus Chair.}
}

\maketitle

\input{00_Abstract}

\begin{IEEEkeywords}
Mobility management, cellular networks, Bayesian optimization, reinforcement learning, data-driven optimization.
\end{IEEEkeywords}

\input{01_Intro}

\input{02_System_model}
\input{03_HDBO}
\input{04_RL}
\input{05_Conclusion}

\bibliographystyle{IEEEtran}
\bibliography{journalAbbreviations, main}

\end{document}

%% file: commands.tex
\def\nb0{{\mathbf{0}}}
\def\nb1{{\mathbf{1}}}

\def\ncalB{{\mathcal{B}}}

\def\sinr{\mathtt{SINR}}			

\def\calB{\mathcal{B}}

%% file: 00_Abstract.tex
\begin{abstract}
Mobility management in cellular networks faces increasing complexity due to network densification and heterogeneous user mobility characteristics. Traditional handover (HO) mechanisms, which rely on predefined parameters such as A3-offset and time-to-trigger (TTT), often fail to optimize mobility performance across varying speeds and deployment conditions. Fixed A3-offset and TTT configurations either delay HOs, increasing radio link failures (RLFs), or accelerate them, leading to excessive ping-pong effects. To address these challenges, we propose two data-driven mobility management approaches leveraging high-dimensional Bayesian optimization (HD-BO) and deep reinforcement learning (DRL). HD-BO optimizes HO parameters such as A3-offset and TTT, striking a desired trade-off between ping-pongs vs. RLF. DRL provides a non-parameter-based approach, allowing an agent to select serving cells based on real-time network conditions. We validate our approach using a real-world cellular deployment scenario, and employing Sionna ray tracing for site-specific channel propagation modeling. Results show that both HD-BO and DRL outperform 3GPP set-1 (TTT of 480 ms and A3-offset of 3 dB) and set-5 (TTT of 40 ms and A3-offset of -1 dB) benchmarks. We augment HD-BO with transfer learning so it can generalize across a range of user speeds. Applying the same transfer-learning strategy to the DRL method reduces its training time by a factor of 2.5 while preserving optimal HO performance, showing that it adapts efficiently to the mobility of aerial users such as UAVs. Simulations further reveal that HD-BO remains more sample-efficient than DRL, making it more suitable for scenarios with limited training data.
\end{abstract}

%% file: 01_Intro.tex
\section{Introduction}
\label{sec:intro}

\subsection{Background and Motivation}
To accommodate growing mobile data traffic and new use cases, network operators are deploying additional infrastructure to enhance coverage, improve spectrum efficiency, and increase spatial reuse \cite{LopPioGer2025,OugGerPol2024}. Managing user mobility in cellular networks remains a critical challenge, particularly with the increasing densification of cells in next-generation networks \cite{shahid2024incorporating, ge2016user}.  

Handover (HO) mechanisms allow a moving user equipment (UE) to seamlessly transition from a serving cell to a target cell while maintaining quality of service (QoS). However, as cellular networks become more dense, frequent HOs can significantly increase the complexity of mobility management \cite{nguyen2020geometry}. The effectiveness of mobility management is heavily influenced by the configuration of the hysteresis margin (A3-offset) and time-to-trigger (TTT)---two parameters playing a crucial role in minimizing ping-pongs and HO failures (HOFs). In conventional cellular networks, UEs typically operate with a finite set of such parameters, which may be adjusted on a per-cell basis. However, due to the interplay between cells, optimizing HO settings individually does not guarantee optimal mobility performance at the network level. Achieving an efficient configuration requires a joint optimization approach, which becomes increasingly complex in large-scale deployments.

Moreover, mobility management must account for UE speed variations, as a parameter set optimized for one speed may not be suitable for another. For instance, high-mobility UEs may travel deep inside a target cell before the TTT expires, increasing the likelihood of HOF due to degraded signal-to-interference-plus-noise ratio (SINR). Conversely, these UEs may also experience unnecessary HOs (ping-pongs) when passing through small cells too quickly \cite{gures2020comprehensive}. 
These challenges are intensified for aerial UEs, such as uncrewed aerial vehicles (UAVs) or drones,
%
%
as they experience rapid fluctuations in received signal strength and strong interference from neighboring cells, further worsening connectivity issues\cite{GerGarAza2022, benzaghta2023designing}. These factors collectively contribute to frequent and unnecessary HOs (ping-pongs), increasing both signaling overhead and the likelihood of radio link failures (RLFs). 

These challenges highlight the need for UE-specific and site-specific solutions to mobility management, beyond a traditional 3GPP approach that relies on a specific set of HO parameters across all network cells. 
To address this, we leverage recent advancements in data-driven models and study the advantages of utilizing either high-dimensional Bayesian optimization (HD-BO) or deep reinforcement learning (DRL) to solve the mobility management optimization problem for real-world cellular network deployments. 

\subsection{Related Work}

Due to its critical importance, mobility management challenges in cellular networks have attracted significant attention from the wireless industry, research community, and standardization bodies. In this section, we provide an overview of the literature on mobility management for both ground UEs (GUEs) and UAVs.  

\subsubsection*{Mobility management for GUEs}

In \cite{shahid2024incorporating}, 
the authors propose a HO algorithm for small cells that integrates UE mobility direction to reduce frequent HOs. 
By analyzing the reference signal received power (RSRP) patterns during the TTT, 
the algorithm detects movement changes, 
allowing HO decisions based on whether the UE is approaching or moving away from a target cell. 
The authors in \cite{karmakar2022mobility} propose an online learning-based mobility management mechanism that computes posterior probabilities of RSRP values to identify the optimal target cell for HO. 
A mathematical framework for a hybrid radio frequency–visible light communication network is presented in \cite{arshad2021stochastic}. 
The study derives UE-to-BS association probabilities and HO rates using stochastic geometry for analytical modeling.
Similarly, \cite{malik2024performance} presents a mathematical analysis for HO triggering estimation based on speed, distance, and predefined mobility models. 
A federated learning-based HO algorithm is proposed in \cite{chien2024privacy}, 
which dynamically adjusts HO thresholds based on predicted RSRPs and historical HO performance to adapt to real-time conditions. 

Several works studied the use of RL in mobility management focusing on non-parameter-based methods,
where the agents make HO decisions by selecting the next serving cell without relying on predefined parameters. 
This way,
the agents learn HO strategies based on real-time network conditions,
leading to more adaptive and dynamic mobility management compared to threshold-based approaches. 
For instance,
a multi-agent Q-learning scheme is proposed for real-time UE association in dense mmWave networks in \cite{Alizadeh2024}. 
The action selection policies are designed to make HO decisions,
which maximize the network sum rate while satisfying load-balancing constraints. 
In \cite{Huynh2021},
a beam association strategy based on Q-learning is designed for high-mobility vehicular networks,
where the RL agent decides whether a vehicle remains connected to the current beam or associates to another beam while also maximizing the overall data rate.
The authors formulate the problem as a semi-Markov decision process and propose a parallel Q-learning approach to deal with multiple vehicles. 
Meanwhile, the authors of \cite{prado2023enabling} focus on proportional fairness in UEs data rates while minimizing the number of HOs. 
They propose single- and multi-agent approaches based on Deep Q-Network (DQN) to select the next serving BS for each UE. 
Moreover, the authors of \cite{dai2024intelligent} introduce a HO management scheme that integrates DRL with reservoir computing in Open-RAN architecture. 
They employ the Proximal Policy Optimization (PPO) algorithm to maximize a weighted sum of Reference Signal Received Quality (RSRQ), UE throughput, and spectral efficiency. 

\subsubsection*{Mobility management for UAVs}

Similar to GUE mobility management, several studies have aimed to improve UAV HO performance. In \cite{Chen2020} and \cite{chen2020deep}, the authors employ Q-learning and DQN, respectively, to dynamically make HO decisions for UAVs. They consider a weighted sum of HOs and the RSRPs of the serving cells as the reward function. 
In \cite{Meer2024}, the researchers propose a mobility management framework to reduce the number of HOs while maintaining service availability. They compare a threshold-based mobility robustness optimization method and a learning-based DQN algorithm for cellular connected UAVs. 
The authors of \cite{Galkin2022} introduce a UAV UE association framework combining deep Q-learning with regression machine learning. The regression neural network estimates the interference between the available BSs, while the Dueling Deep Q-Network selects the next serving BS. The proposed solution aims to maximize the overall throughput while minimizing the HO rate.

\bigskip 

Despite notable progress in mobility management for both GUEs and UAVs, the community still faces the challenge of developing a scalable optimization framework that efficiently leverages available data to improve practical mobility management key performance indicators (KPIs). Mathematical models such as those in \cite{arshad2021stochastic, malik2024performance} are analytically insightful but not suited for large-scale, real-world deployments. Studies like \cite{Huynh2021, karmakar2022mobility, Chen2020, chen2020deep} typically address a single UE mobility profile, limiting the generalization capabilities of the proposed frameworks. Others, such as \cite{shahid2024incorporating, chien2024privacy}, require expert knowldge to adjust HO parameters. These limitations hinder their ability to fully exploit available data or to generalize across diverse mobility scenarios. To address these gaps, we propose scalable, data-driven mobility management approaches based on high-dimensional Bayesian optimization (HD-BO) and deep reinforcement learning (DRL). Our study focuses on the trade-offs between ping-pongs vs. HOF and ping-pongs vs. RLF, analyzing performance across varying UE speed categories and demonstrating the practical viability of these approaches in large-scale site-specific cellular network deployments.


%
%

\subsection{Approach and Contribution}

We provide two methodologies based on HD-BO and DRL for scalable cellular mobility management, optimizing practical HO metrics for diverse speeds across designated streets.

Although BO \cite{shahriari2015taking} has proven effective in addressing coverage-capacity tradeoffs and optimizing radio resource allocation \cite{dreifuerst2021optimizing, eller2024differentiable, zhang2023bayesian, maggi2021bayesian, tambovskiy2022cell, maggi2023energy, tekgul2023joint, de2023towards}, it is inherently limited by the number of decision variables it can efficiently handle—typically around twenty or fewer in continuous domains \cite{frazier2018tutorial}. This constraint restricts the scalability of BO for optimizing mobility parameters in large-scale cellular networks. To overcome these limitations, this paper takes the first step in employing high-dimensional Bayesian optimization (HD-BO) to optimize mobility-related HO KPIs. To the best of our knowledge, this paper is the first to (i) apply HD-BO tools to address practical mobility management challenges in large-scale cellular networks using real-world scenarios, and (ii) explore model generalization to diverse UE speeds within the context of transfer learning through HD-BO.

In addition, we introduce a non-parameter-based model-free mobility management approach leveraging DRL. Unlike the HD-BO method, which requires predefined parameters for HO decisions, the DRL-based solution enables an agent to dynamically select the next serving cell for a certain UE based on network state information, eliminating the need for static thresholds.

Our main contributions can be summarized as follows:
\begin{itemize}[leftmargin=*]

\item
\emph{High-dimensional BO for HO parameter-based mobility management:}
We apply a state-of-the-art HD-BO technique to mobility management, demonstrating its effectiveness in optimizing HO decisions for both GUEs and UAVs. Specifically, we identify optimal A3-offset and TTT configurations for real-world cellular network deployments, that balance conflicting HO KPIs, such as ping-pongs vs. HOF and ping-pongs vs. RLF. By extending BO to high-dimensional settings, we enable large-scale, data-driven mobility optimization beyond traditional BO constraints. Our case studies consider both GUEs and UAVs moving at varying speeds. 
Our extensive evaluations on a real-world cellular deployment scenario in London show that,
for per-cell optimization, 
HD-BO reduces ping-pongs by 73\% for GUEs moving at 60 km/h compared to 3GPP benchmarks. 
Also, for UAVs at a 150\,m altitude, 
HD-BO outperforms 3GPP set-1 and set-5 benchmarks, 
achieving a 3\% ping-pong rate (vs. 15\% and 11\%) and 0\% RLF similar to the upper-bound set-5, and performing better than set-1 of 9\% RLF.

\item
\emph{DRL for parameter-free mobility management:}  
We compare the performance of the HD-BO approach to a non-threshold-based mobility management method utilizing DRL. Unlike the HD-BO method, which optimizes predefined HO parameters such as A3-offset and TTT, the DRL-based solution eliminates the reliance on fixed HO thresholds. Instead, an agent autonomously learns and selects the optimal serving cell in real-time based on the network state by directly interacting with the environment. 
Our results show that the performance achieved through DRL is comparable to HD-BO in both ping-pong reduction and RLF minimization, 
confirming that DRL is a viable alternative to parameter-based mobility management,
without predefined A3-offset and TTT thresholds.

\item
\emph{Transfer learning:} Aiming at faster convergence to optimal solutions, and aligning with the 3GPP vision on the need for data-driven model generalization \cite{3GPP38.843}, we explore the \emph{transfer learning} capabilities of the HD-BO and DRL approaches. 
We use transfer learning to leverage measurement outcomes from a previously performed optimization process, denoted as the \emph{scenario source}, to predict the best solution for a new optimization, termed the \emph{scenario target}. Our experiments reveal that the HD-BO approach is capable of generalization to diverse UE speeds; furthermore, transfer learning applied through DRL reduces training time by 2.5× while maintaining optimal HO performance.
\end{itemize}

%% file: 02_System_model.tex
\section{System Model}
\label{sec:system_model}
In this section, we describe the network deployment, channel model, and mobility management performance metrics used in our study.  

\subsection{Cellular Topology and Site-specific Propagation Channel} 

We consider a site-specific scenario based on a real-world production radio network operated by a leading commercial mobile provider in the UK. 

\subsubsection*{Cellular network deployment}
The deployment under study consists of 10 cell sites, with antenna heights ranging from 22\,m to 56\,m. Each site comprises three sector antennas, resulting in a total of 30 cells across the network. The selected geographical area spans $1400 \times 1275$\,m and is located in London, between latitudes $[51.5087, 51.5215]$ and longitudes $[-0.1483, -0.1296]$.  
A 3D representation of the selected area is constructed using OpenStreetMap and Blender, incorporating both terrain and building information. In our first case study, focusing on ground UE mobility management, the BSs antennas are configured according to the actual cellular network. In our second case study, focusing on UAV mobility management, BS antennas are optimized to achieve a trade-off between ground and aerial coverage. More details are provided in Section~\ref{sec:case_study_2}.

\subsubsection*{Propagation channel}
The channel between BS $b$ and UE $k$ is computed using Sionna RT \cite{hoydis2023sionna}, a widely used 3D ray-tracing tool for site-specific radio wave propagation analysis. Simulations are performed at a carrier frequency of $2$\,GHz. The material \texttt{itu_concrete} is used to model the permittivity and conductivity of all buildings. The maximum number of reflections and diffractions is set to 5 and 1, respectively.  

\subsubsection*{SINR formulation}
We compute the downlink wideband SINR in dB experienced by UE $k$ from its serving BS $b_k$ as
\begin{equation}
  \sinr_{\textrm{dB},k} = 10\,\log_{10} \left( \,\frac{p_{b_k} \cdot G_{b_k,k}}{
  \sum\limits_{b\in\calB\backslash b_k}{p_{b} \cdot G_{b,k}  \,+\, \sigma_{\textrm{T}}^2}}\right),
  \label{SINR_DL_TN}
\end{equation}
where $G_{b,k}$ is the square magnitude of the channel gain, incorporating both small-scale and large-scale fading, averaged over 50 physical resource blocks (PRBs), each with a bandwidth of 180\,kHz. The thermal noise power $\sigma_{\textrm{T}}^2$ over 10\,MHz is obtained from a power spectral density of $-174$\,dBm/Hz, while the transmit power of BS $b$ across the entire bandwidth is $p_{b} = 46$\,dBm \cite{3GPP36814}.

\subsubsection*{Cellular mobility management problem}
\begin{figure}
\centering
\includegraphics[width=\figwidth]{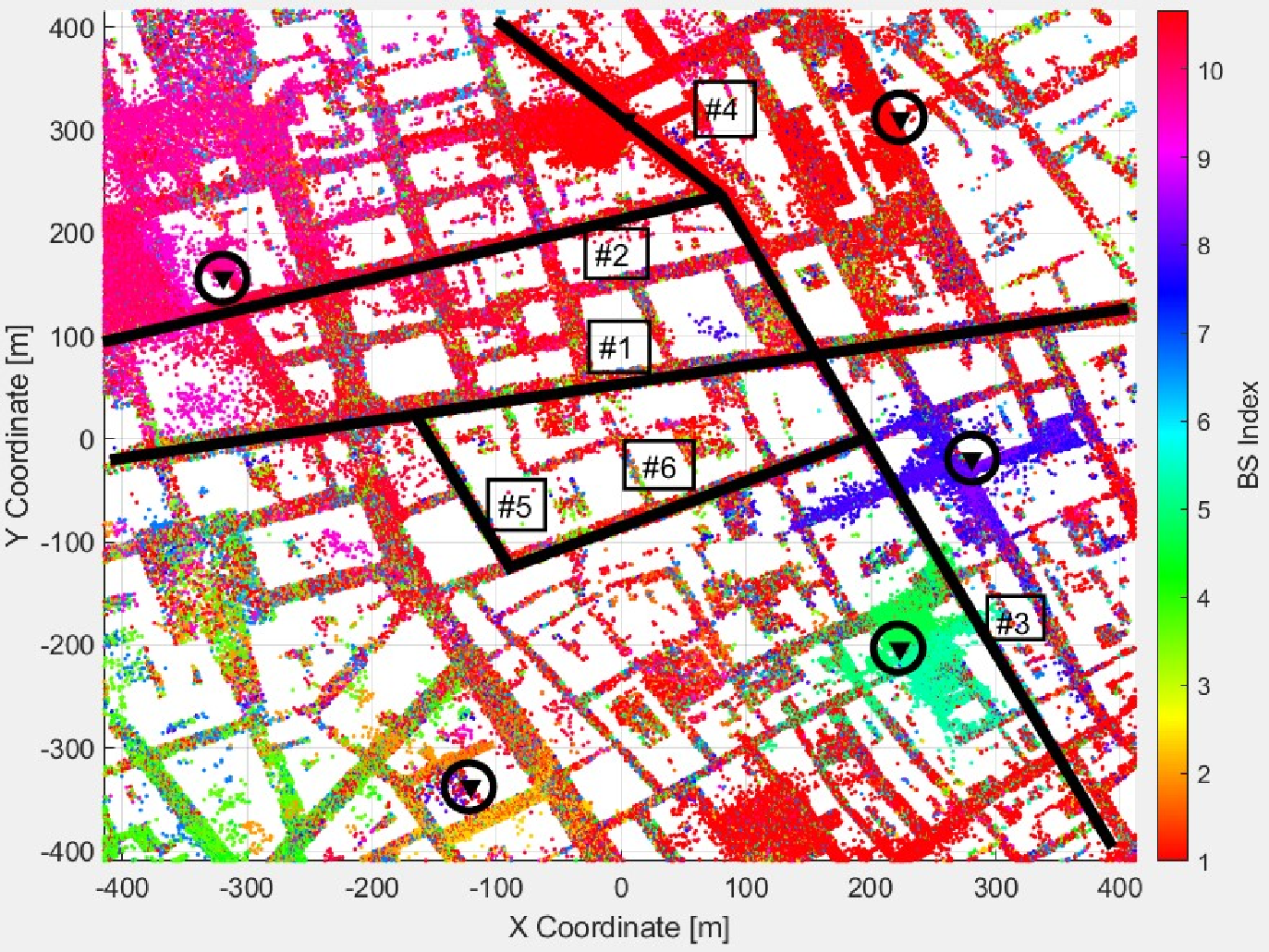}
\caption{2D representation of the selected urban area, illustrating UE positions along streets (colored dots) and cell deployment sites (triangles). The five chosen streets are marked with black lines, and colors denote the index of the cell site providing the strongest average received power.}
\label{fig:streets}
\end{figure}

For our mobility study, we choose five main streets within the selected geographical area of London for experimentation. Fig.~\ref{fig:streets} provides a 2D representation of the selected urban area, where colored dots indicate outdoor locations of UEs along the streets. The five selected streets are marked with black lines, with their corresponding IDs labeled beside them. Some of the BS deployment sites are represented as circled triangles, where each site consists of three sector antennas. The color of each dot indicates the cell site (one out of ten) providing the strongest average received power, as shown by the heatmap bar on the side. This visualization highlights the challenges of mobility management in real-world scenarios, since rapid signal fluctuations may cause frequent changes in serving cells (handovers) over short distances. Consequently, an effective data-driven mobility management framework is required, one that is robust and capable of targeting site-specific scenarios.  

\subsection{Handover Mechanism in Cellular Networks} 

In the following, we further detail the cellular mobility management problem and introduce the concepts of HO, HOF, RLF, and ping-pongs.  

\subsubsection*{Handover process} 
In cellular networks, handovers can occur between different radio access technologies (RATs), carriers, or cells \cite{3GPP36.839}. In this study, we focus on intra-RAT, intra-carrier HOs and on hard HOs, where a UE disconnects from the source cell before establishing a new connection with the target cell.%
\footnote{Handovers can be governed by both signal strength and signal quality. In this study, we follow the 3GPP assumptions in \cite{3GPP36.839}, where the handover procedure is based on Reference Signal Received Power (RSRP) measurements.}
\subsubsection*{Handover measurements} 
The UE performs HO measurements and processing at both Layer 1 (physical layer) and Layer 3 (network layer). For HO measurements, the UE typically estimates the RSRP for the cells listed in its neighboring cell list. To mitigate the effects of small-scale fading in RSRP estimations, the UE computes each RSRP sample as the linear average of power contributions from all resource elements carrying reference symbols within a single subframe and the designated measurement bandwidth (e.g., six PRBs). These averaged RSRP samples are then further smoothed over multiple samples. This linear averaging process, performed at Layer 1, is known as \emph{L1 filtering}. In a typical setup, downlink RSRP samples may be collected every 40 ms and then averaged over five successive samples to obtain an L1-filtered HO measurement \cite{lopez2012mobility}. The L1-filtered HO measurements are further averaged using a first-order infinite impulse response filter to mitigate the effects of fading and estimation imperfections. This moving averaging process is performed at Layer 3 and is referred to as \emph{L3 filtering}. A typical L3 filtering period is 200 ms. For further details on the L1 and L3 filtering procedures, we refer the reader to Fig.~1 in \cite{lopez2012mobility}. 

\subsubsection*{Handover trigger} 
A HO is triggered when the L3-filtered HO measurement satisfies a HO event entry condition. While there are eight types of HO event entry conditions \cite[Section 5.5.4]{3GPP36.331}, we focus on \emph{Event A3: `neighbor becomes offset better than server'}, since it is typically use to trigger intra-RAT intra-carrier HOs \cite{lopez2012mobility}. Once the A3 condition is met---i.e., the L3-filtered RSRP of the target cell exceeds that of the serving cell by a hysteresis margin (also known as the event \emph{A3 offset})---the UE initiates a \emph{Time-to-Trigger (TTT)} timer. The UE initiates the HO preparation process only if the event A3 condition remains satisfied throughout the TTT period. If that is the case, the UE notifies the serving cell and reports the event A3 condition via a measurement report. Then, the HO preparation phase begins.

\subsubsection*{Handover preparation and execution} 
The source cell issues a HO request message to the target cell, which performs admission control procedures based on the quality of service requirements of the UE. The target cell prepares for the HO process and sends a HO request acknowledgment to the source cell.
Upon receiving the HO request acknowledgment, the source cell initiates data forwarding to the target cell and sends a HO command to the UE. Finally, in the HO execution phase, the UE synchronizes with the target cell and establishes access. Once the HO procedure is completed, the UE sends a HO complete message to the target cell, allowing the target cell to begin data transmission to the UE \cite{3GPP36.839, lopez2012mobility}.  

\subsubsection*{Radio link failure and handover failure}
RLF occurs when a UE is unable to maintain a reliable connection with the serving cell due to sustained poor signal quality. Specifically, a UE is considered out of synchronization when its wideband SINR, denoted by $\sinr_{\textrm{dB},k}$, falls below a threshold $Q_{\text{out}}$. The UE regains synchronization when this SINR exceeds a threshold $Q_{\text{in}}$. Once $\sinr_{\textrm{dB},k}$ drops below $Q_{\text{out}}$, a timer $T_{310}$ is triggered. If the SINR does not recover above $Q_{\text{in}}$ before $T_{310}$ expires, the UE declares a RLF \cite{3GPP36.839, 3GPP36.300}. HOF is then defined as a specific instance of RLF occurring during the HO process. Based on \cite{3GPP36.839, 3GPP36.300}, a HOF is declared if any of the following conditions are met: (1) RLF occurs after the HO is triggered (e.g., event A3) but before the HO command is received; (2) the $T_{310}$ timer is already running when the handover command is sent; or (3) $\sinr_{\textrm{dB},k}$ remains below $Q_{\text{out}}$ at the time the handover complete message is sent. In these cases, the HO cannot be successfully completed due to poor radio conditions. For the sake of tractability, in our model, we adopt the definition of HOF as follow: a HO is considered to have failed if the UE’s SINR is below $Q_{\text{out}}$ at the moment the HO complete message is sent. This simplification allows us to assess HOF events directly based on instantaneous SINR measurements during the handover execution phase.

\subsubsection*{Handover ping-pongs}
The occurrence of a HO ping-pong is determined by the duration for which a UE remains connected to a cell immediately after a handover, referred to as the \emph{time-of-stay}. This duration begins when the UE sends a handover complete message to the target cell and ends when the UE sends another HO complete message to a new cell. A HO is classified as a ping-pong if the UE’s time-of-stay is shorter than a predefined threshold, $T_p$ (e.g., 1\,s), and if the new target cell is the same as the original source cell prior to the previous HO, leading to increased signaling overhead and reduced network efficiency.

\subsubsection*{Performance trade-off}

Small A3 offsets and TTT values can trigger premature handovers, increasing the ping-pong effect, while larger values may excessively delay handovers, leading to a higher risk of HOF or RLF. Therefore, optimizing the A3 offsets and TTT based on UE velocity and site-specific radio propagation conditions is crucial for effective mobility management. 
In 3GPP, mobility management relies on predefined threshold sets to regulate handover decisions. Among the benchmark configuration suggested by the 3GPP in the simulation recommendations, we specifically focus on \emph{set-1} and \emph{set-5} \cite{3GPP36.839}, as they represent two extreme cases in handover optimization:
\begin{itemize}
\item 
Set-1 is designed to reduce ping-pongs by delaying handovers, applying a uniform configuration across all cells with a TTT of 480\,ms and an A3-offset of 3\,dB. \item 
Set-5, on the other hand, aims to minimize HOF by accelerating handovers, setting TTT to 40\,ms and A3-offset to -1\,dB, again uniformly across all cells. While this configuration allows the UE to switch to a stronger serving cell more quickly, it increases the likelihood of ping-pongs, as rapid transitions may cause frequent unnecessary handovers. \end{itemize}
We employ 3GPP set-1 and set-5 as performance benchmarks, as they define the two extremes in handover performance trade-offs, serving as a reference for evaluating the performance of our data-driven methods. It should be noted that 3GPP evaluations apply these settings uniformly across all cells in their performance evaluations rather than adapting them per-cell, primarily to simplify network-wide mobility management and reduce optimization complexity. However, this approach is often suboptimal, as it lacks adaptability to site-specific radio conditions and UE mobility patterns, motivating the need for adaptive, data-driven optimization techniques.

In the remainder of this paper, we demonstrate how data-driven machine learning methods can optimize mobility management by addressing two key optimization problems: i) balancing ping-pongs and HOF, ii) balancing ping-pongs and RLF. The exact problem formulations will be detailed in the following two sections, where we introduce two different optimization methodologies based on HD-BO and DRL, respectively.

%% file: 03_HDBO.tex
\section{Cellular Mobility Management via\\High-dimensional Bayesian Optimization}
\label{sec:BO}

In this section, we formulate the HO parameter-based mobility management optimization problem and propose a solution based on high-dimensional Bayesian optimization (HD-BO). 

\subsection{Problem Formulation}

Our objective is to identify the joint optimal sets of A3-offset and TTT parameters for all cells under consideration that minimize conflicting HO KPIs. We consider two practical trade-offs defined as follows.

\emph{KPI study \#1: Ping-pongs vs. HOF.}
We examine the trade-off between reducing the number of ping-pongs and reducing the number of HOF. The problem is formally defined as follows: 
\begin{align}
\min_{\text{\textbf{A3}},\text{\textbf{TTT}}} \;\; w_{\text{PP}} \cdot \frac{\sum_{t} \mathds{1}_{\text{PP}_t}}{\sum_{t} \mathds{1}_{\text{HO}_t} - \mathds{1}_{\text{HOF}_t}} + w_{\text{HOF}} \cdot \frac{\sum_{t} \mathds{1}_{\text{HOF}_t}}{\sum_{t} \mathds{1}_{\text{HO}_t}},
\label{eqn:Opt_problem_joint_HOF} 
\end{align} 
\begin{align}
\text{s.t.} \quad & \text{A3}_b \in \left( \underline{\text{A3}}, \overline{\text{A3}} \right), \enspace b = 1, \ldots, \ncalB \tag{2a} \\
& \text{TTT}_{b} \in \left( \underline{\text{TTT}}, \overline{\text{TTT}} \right), \enspace b = 1, \ldots, \ncalB \tag{2b}
\end{align}
where $\mathds{1}_{(\cdot)}$ is an indicator function that evaluates whether a HO, a ping-pong, or a HOF event occurs at time-step $t$. This function take a value of 1 if the respective event is observed and 0 otherwise. 
The objective function (\ref{eqn:Opt_problem_joint_HOF}) minimizes a weighted sum, where $w_{\text{PP}}$ and $w_{\text{HOF}}$ are positive real numbers that determine the relative importance of reducing ping-pongs versus HOF. The vectors $\text{\textbf{A3}}$ and $\text{\textbf{TTT}}$ contain the A3-offset $\text{A3}_b$ and time-to-trigger $\text{TTT}_b$ of all BSs $b$ $\in$ $\calB$, respectively. The smallest allowed values are $\underline{\text{A3}}$ and $\underline{\text{TTT}}$, while ${\overline{\text{A3}}}$, ${\overline{\text{TTT}}}$ are the largest allowed values.

\emph{KPI study \#2: Ping-pongs vs. RLF.}
In this experiment, we focus on RLF instead of HOF, i.e., on reducing link outages while minimizing the number of ping-pongs.
Similarly, the problem is formally defined as follows:

\begin{align}
\min_{\text{\textbf{A3}},\text{\textbf{TTT}}} \;\; w_{\text{PP}} \cdot \frac{\sum_{t} \mathds{1}_{\text{PP}_t}}{\sum_{t} \mathds{1}_{\text{HO}_t} - \mathds{1}_{\text{HOF}_t}} + w_{\text{RLF}} \cdot \frac{\sum_{t} \mathds{1}_{\text{RLF}_t}}{\sum_{t} \mathds{1}_{\text{HO}_t}},
\label{eqn:Opt_problem_joint_RLF} 
\end{align}   
\begin{align}
\text{s.t.} \quad & \text{A3}_b \in \left( \underline{\text{A3}}, \overline{\text{A3}} \right), \enspace b = 1, \ldots, \ncalB \tag{3a} \\
& \text{TTT}_{b} \in \left( \underline{\text{TTT}}, \overline{\text{TTT}} \right), \enspace b = 1, \ldots, \ncalB \tag{3b}
\end{align}

The optimization problems (\ref{eqn:Opt_problem_joint_HOF}) and (\ref{eqn:Opt_problem_joint_RLF}) are nonconvex due to the nonconcavity of the objective functions. Furthermore, for our practical scenario with 30 cells, they involve 60 optimization variables, requiring efficient methods to handle the large search space.
In the following, we introduce high-dimensional Bayesian optimization as a potential solution for optimizing HO parameter-based mobility management.

\subsection{High-dimensional Bayesian Optimization}

Bayesian optimization (BO) operates by iteratively constructing a probabilistic \textit{surrogate model} of the objective function 
$f(\cdot)$ based on prior evaluations at selected points \cite{shahriari2015taking}. This surrogate model, which is computationally easier to evaluate than $f(\cdot)$, is continuously updated as new points are assessed. To determine the next point to evaluate, an acquisition function $\alpha(\cdot)$ scores the surrogate model’s response, guiding the search process. The acquisition function strategically balances exploration (searching for new, potentially better solutions) and exploitation (refining the current best solutions).

\subsubsection*{Objective function evaluation}
We define a query point $\textbf{x} = [\textbf{A3}, \textbf{TTT}]$ as a configuration for both the A3-offset and TTT for each BS $b \in \mathcal{B}$, and obtain the corresponding objective function value $f(\textbf{x})$ as in (\ref{eqn:Opt_problem_joint_HOF}) or (\ref{eqn:Opt_problem_joint_RLF}). In both cases, the objective function $f(\cdot)$ being optimized is a mathematically intractable stochastic function that captures the model detailed in Section~\ref{sec:system_model}, along with the inherent randomness of UE locations and the wireless channel. As a result, we do not directly observe $f(\textbf{x})$; instead, we obtain a noisy realization or observation of the function, denoted by $\tilde{f}(\textbf{x})$, through system-level simulations. Importantly, repeated evaluations at the same query point $\textbf{x}$ may yield different outcomes due to the intrinsic stochasticity of the environment. For convenience, we define a set of $N$ query points as $\textbf{X} = [\textbf{x}_1, \ldots, \textbf{x}_N]$ and the corresponding set of observations as $\tilde{\textbf{f}}(\textbf{X}) = [\tilde{f}_1, \ldots, \tilde{f}_N]^\top$, where each $\tilde{f}_i = \tilde{f}(\textbf{x}_i)$ represents a stochastic observation of $f(\textbf{x}_i)$ for $i = 1, \ldots, N$. In practical deployments, these observations could also be obtained through real-world network measurements.

\subsubsection*{Gaussian Process prior distribution}
We use a Gaussian Process (GP) prior, $\widehat{f}(\cdot)$, to construct a surrogate model (i.e., the posterior) that approximates the objective function $f(\cdot)$ \cite{shahriari2015taking}. The resulting GP model enables the prediction of $\tilde{f}(\textbf{x})$ at a query point $\textbf{x}$ based on previously observed values, $\tilde{\textbf{f}}(\textbf{X})=\tilde{\textbf{f}}$, over which the model is trained. 
Formally, the GP prior on the objective function $\tilde{f}(\textbf{x})$ assumes that for any set of input points $\textbf{X}$, the corresponding function values $\tilde{\textbf{f}}$ are jointly distributed as  

\begin{equation}
  p(\,\tilde{\textbf{f}}\,) = \mathcal{N}(\,\tilde{\textbf{f}} \,\,|\,\, \boldsymbol{\mu}(\textbf{X}),\mathbf{K}(\textbf{X})\,),
  \label{posterior}
\end{equation}
where $\boldsymbol{\mu}(\textbf{X}) = [\mu(\mathrm{\textbf{x}}_1),\ldots,\mu(\mathrm{\textbf{x}}_N)]^\top$ is the $N \times 1$ mean vector,  
and $\mathbf{K}(\textbf{X})$ is the $N \times N$ covariance matrix,  
with each entry $(i,j)$ given by the covariance function $k(\textbf{x}_{i},\textbf{x}_{j})$.  
For a given point $\textbf{x}$, the mean function $\mu(\textbf{x})$ provides prior knowledge about $f(\textbf{x})$, while the kernel function $\mathbf{K}(\textbf{X})$ captures the uncertainty between different input values $\textbf{x}$.  

\subsubsection*{Gaussian Process posterior distribution}
Given a set of observed noisy samples $\tilde{\textbf{f}}$ at previously sampled points $\textbf{X}$, the posterior distribution of $\widehat{f}(\textbf{x})$ at a new query point $\textbf{x}$ can be expressed as \cite{frazier2018tutorial}:

\begin{equation}
  p(\widehat{f}(\textbf{x}) = \widehat{f} \,\, | \,\, \textbf{X}, \tilde{\textbf{f}} \,) = \mathcal{N}(\widehat{f} \,\,|\,\, \mu(\textbf{x} \,|\, \textbf{X}, \tilde{\textbf{f}}),\sigma^2(\textbf{x} \,|\, \textbf{X}, \tilde{\textbf{f}})),
  \label{posterior_Noisy}
\end{equation}
where the posterior mean and variance are given by:

\begin{equation}
  \mu(\textbf{x} \,|\, \textbf{X},\tilde{\textbf{f}}) = \mu(\textbf{x}) + \tilde{\textbf{k}}(\textbf{x})^\top (\tilde{\textbf{K}}(\textbf{X}))^{-1}(\tilde{\textbf{f}}-\boldsymbol{\mu}(\textbf{X})),
  \label{Mean_posterior_Noisy}
\end{equation}

\begin{equation}
  \sigma^2(\textbf{x} \,|\, \textbf{X},\tilde{\textbf{f}}) = k(\textbf{x},\textbf{x}) - \tilde{\textbf{k}}(\textbf{x})^\top (\tilde{\textbf{K}}(\textbf{X}))^{-1} \,\tilde{\textbf{k}}(\textbf{x}),
  \label{Kernel_posterior_Noisy}
\end{equation}
where  
$\tilde{\textbf{k}}(\textbf{x}) = [k(\textbf{x},\textbf{x}_{1}),\ldots,k(\textbf{x},\textbf{x}_{N})]^\top$ is the $N \times 1$ covariance vector,  
and $\tilde{\textbf{K}}(\textbf{X}) = \textbf{K}(\textbf{X}) + \sigma^2 \textbf{I}_{N}$,  
with $\sigma^2$ representing the observation noise (i.e., the variance of the Gaussian distribution),  
and $\textbf{I}_{N}$ denoting the $N \times N$ identity matrix.  
Note that \eqref{Mean_posterior_Noisy} and \eqref{Kernel_posterior_Noisy} define the mean and variance of the estimated function $\widehat{f}(\textbf{x})$, where the variance quantifies the uncertainty in the prediction.

\subsubsection*{Initial dataset creation and acquisition function}
The BO algorithm begins by constructing a Gaussian Process (GP) prior $\{\mu(\cdot), k(\cdot, \cdot)\}$ using an initial dataset  
$\mathcal{D} = \{\textbf{x}_1,\ldots,\textbf{x}_{N_{\textrm{o}}},\tilde{f}_1,\ldots,\tilde{f}_{N_{\textrm{o}}}\}$,  
which consists of $N_{\textrm{o}}$ initial observations. The dataset is generated through system-level simulations based on the objective function defined in (\ref{eqn:Opt_problem_joint_HOF}) or (\ref{eqn:Opt_problem_joint_RLF}) and the model described in Section~\ref{sec:system_model}.  
For each observation point $\textbf{x}_i \in \mathcal{D}$, the A3-offset and TTT values are randomly selected from the ranges $[-1\text{ dB}, 3\text{ dB}]$ and $[40 \text{ ms}, 480 \text{ ms}]$, respectively.  
The algorithm then leverages the observations in $\mathcal{D}$ to choose $\textbf{x}_n$. This is performed via an acquisition function $\alpha(\cdot)$, which is designed to trade off the exploration of new points in less favorable regions of the search space with the exploitation of well-performing ones. 
The former prevents getting caught in local minimum, 
whereas the latter minimizes the risk of testing points with excessively degrading performance. We adopt Thompson sampling as the acquisition function, which has shown to perform well in terms of balancing the trade-off between exploration and exploitation \cite{eriksson2019scalable}. 

\subsubsection*{Batch evaluation of candidate points}
We employ a batch evaluation strategy that enables efficient query space exploration while reducing the number of required physical experiments. At each iteration, a set of $N_{\textrm{c}}=500$ candidate points is selected based on the posterior distribution (\ref{posterior_Noisy}) and evaluated in parallel across available computational resources. This approach leverages the capability of BO to learn from a limited number of samples, making it particularly suitable for scenarios where extensive real-world experimentation is impractical. We split the candidate points into 10 batches each consisting of 50 points. The query
point $\textbf{x}_n$ is then chosen as
\begin{equation}
    \textbf{x}_n = \underset{\substack{i}}{\textrm{arg}\,\textrm{max}}\;\; \alpha \left(\textbf{x}_{\text{cand}_i} \,|\, \mathcal{D} \right).
    \label{eqn:BO_acq}
\end{equation} 
Once $\textbf{x}_n$ is determined, a new observation of the objective function $\tilde{f}(\textbf{x}_n)$ is then produced, and the dataset $\mathcal{D}$, the GP prior, and the best observed objective value $\tilde{f}^{*}$ are all updated. 

For BO methods to achieve greater sample efficiency, it is essential to introduce a hierarchical significance structure for the dimensions of $\textbf{x} \in D$. In high-dimensional problems, certain features, such as $\{\textbf{x}_{22}, \textbf{x}_{44}\}$, may play a critical role in capturing the primary variations of the objective function $f$, while others, such as $\{\textbf{x}_{2}, \textbf{x}_{4}, \textbf{x}_{60}\}$, may have moderate significance. The remaining features may contribute negligibly. HD-BO exploits these hierarchical relationships to improve optimization efficiency. In the following, we introduce the core features of a HD-BO method known as Trust Region BO (TuRBO) \cite{eriksson2019scalable}.%
\footnote{We implemented and tested three HD-BO methods: Sparse Axis-Aligned Subspaces (SAASBO) \cite[Section~4]{eriksson2021high}, BO via Variable Selection (VSBO) \cite[Section~3]{shen2021computationally}, and Trust Region BO (TuRBO) \cite[Section~2]{eriksson2019scalable}. TuRBO demonstrated superior performance and higher suitability for the problem under consideration.}

\subsubsection*{Trust Region BO (TuRBO)}
To address the challenges of high dimensionality in BO, the authors in \cite{eriksson2019scalable} proposed Trust Region BO (TuRBO), an approach that shifts from global surrogate modeling to managing multiple independent local models, each focusing on a distinct region of the search space. TuRBO achieves global optimization by simultaneously maintaining several local models and allocating samples using an implicit multi-armed bandit strategy. This enhances the acquisition strategy by directing samples toward promising local optimization efforts. TuRBO leverages trust region (TR) methods from stochastic optimization, which are gradient-free and employ a simple surrogate model within a defined TR—typically a sphere or polytope centered around the best solution found. However, simple surrogate models may require excessively small trust regions for accurate modeling. To mitigate this, TuRBO utilizes a GP surrogate model within the TR, preserving key BO features such as noise robustness and systematic uncertainty handling. In TuRBO, the TR is defined as a hyperrectangle centered at the current optimal solution, $f^{*}$. The initial side length of the TR is set to $L \gets L_{\text{init}}$. Each dimension's side length is then adjusted according to its respective length scale $\lambda_i$ in the GP model. The side length for each dimension is given by:
\begin{equation}
L_i = {\lambda_i L}\cdot{\left(\sideset{}{_{j=1}^d}\prod \lambda_j \right)^{-1/d}}\!\!\!\!\!\!.
\end{equation}
where $d$ is the total number of dimensions (i.e, optimization parameters under consideration). During each local optimization run, an acquisition function selects a batch of $q$ candidates at each iteration, ensuring they remain within the designated TR. If the TR’s side length $L$ was large enough to cover the entire search space, this method would be equivalent to standard vanilla-BO. Thus, adjusting $L$ is crucial: the TR must be large enough to encompass promising solutions while remaining compact enough to ensure the local model's accuracy.  
The TR is dynamically resized based on optimization progress: it is doubled ($L \gets \min \{L_{\text{max}}, 2L\}$) after $\tau_{\text{succ}}$ consecutive successes and halved ($L \gets L/2$) after $\tau_{\text{fail}}$ consecutive failures. A success is defined as an iteration where the objective function value improves compared to the previous one, whereas a failure corresponds to an iteration with no improvement.
Success and failure counters are reset after each adjustment. If $L$ falls below $L_{\text{min}}$, the TR is discarded, and a new one is initialized at $L_{\text{init}}$. The TR’s side length is capped at $L_{\text{max}}$.  
TuRBO maintains $m$ trust regions simultaneously, denoted as $\text{TR}_{l}$, where $l \in \{1, \dots, m\}$, each defined as a hyperrectangle with a base side length $L_{l} \leq L_{\text{max}}$. Candidate selection involves choosing a batch of $q$ candidates from the union of all TRs. Thompson sampling is used for selecting candidates both within and across trust regions.  

In this study, TuRBO is run using an open-source repository \cite{eriksson2019scalable} with the following hyperparameters: $\tau_{\text{succ}} = 3, \tau_{\text{fail}} = 15, L_{\text{init}} = 0.8, L_{\text{min}} = 2^{-7}, L_{\text{max}} = 1.6$.


\subsection{Case Study \#1: Ground UE Mobility Management}

In our first case study, we analyze GUE mobility across three different speed categories: i) pedestrian speed of 3\,km/h, ii) moderate speed of 30\,km/h, iii) high speed of 60\,km/h. Our study aims to understand the cross-impact of speed-specific optimizations, examining how optimizing HO parameters for one speed affects the performance of others. Additionally, we evaluate the capability of HD-BO to handle heterogeneous speed scenarios, where GUEs move along a street portfolio consisting of five main roads in London. Finally, we compare the benefits of per-cell optimization versus applying a uniform TTT and A3-offset across the entire network, highlighting the advantages of cell-specific mobility management.



\subsubsection*{Minimizing HOF vs. RLF}

Table~\ref{tab:KPI_performance_compare} compares the performance of GUEs for a speed of 3km/h for the two KPI trade-offs:
\begin{itemize}
\item
Ping-pongs vs. HOF (`PP-HOF', KPI study \#1), i.e., problem (2) with $w_{\text{PP}} = 1$ and $w_{\text{HOF}} = 9$.
\item 
Ping-pongs vs. RLF (`PP-RLF', KPI study \#2), i.e., problem (3) with $w_{\text{PP}} = 1$ and $w_{\text{RLF}} = 9$.
\end{itemize}
These optimizations are achieved through the per-cell joint tuning of the A3-offset and TTT parameters using HD-BO and are compared to the reference performances under 3GPP benchmark configurations set-1 and set-5. For our evaluations, we set $Q_{\text{out}} = -8$\,dB,  $T_{310}$ = 1\,s, and $T_p$ = 1\,s, as per \cite{3GPP36.839}.

\begin{figure}
\centering
\includegraphics[width=\figwidth]{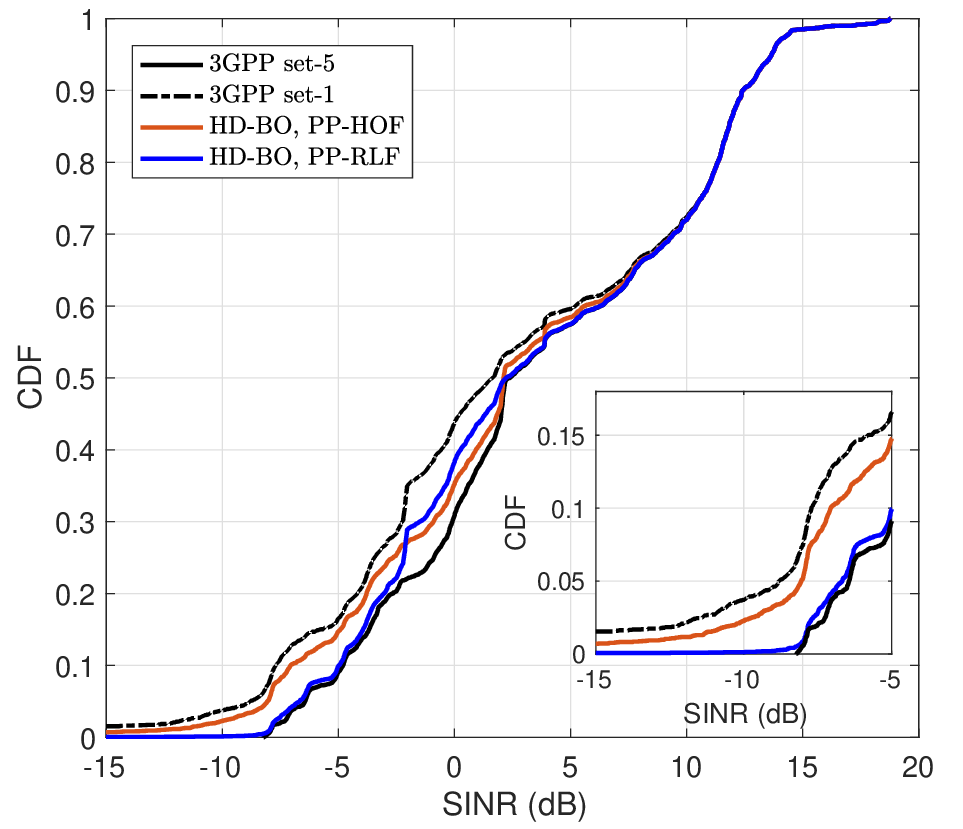}
\caption{SINR for GUEs at 3\,km/h: the 3GPP benchmark configurations (set-1 and set-5), optimizing ping-pongs vs. HOF (`PP-HOF'), and optimizing ping-pongs vs. RLF (`PP-RLF').}
\label{fig:KPI_compare_HOF_RLF}
\end{figure}

Fig.~\ref{fig:KPI_compare_HOF_RLF} illustrates the cumulative distribution function (CDF) of the SINR for UEs of speed 3\,km/h. The solid and dashed black lines represent the SINR performance under the 3GPP benchmark configurations, set-5 and set-1, respectively. The green curve in Fig.~\ref{fig:KPI_compare_HOF_RLF} represents the performance after the data-driven optimization for KPI study \#1, while the blue curves depict the performance for KPI study \#2. Based on the results in Table~\ref{tab:KPI_performance_compare} and Fig.~\ref{fig:KPI_compare_HOF_RLF}, the following key observations can be drawn:

\begin{itemize}[leftmargin=*]
\item
The HD-BO approach successfully reduces HOF to 0\% in KPI study \#1 and RLF to 0\% in KPI study \#2, achieving the upper bound set by 3GPP set-5 while also reducing ping-pongs by 25\%.

\item 
Minimizing RLF
leads to better outage performance than minimizing HOF. Specifically, when focusing on HOF reduction, the optimization framework may delay handovers to avoid reporting HOF at the measurement time. While this results in a 0\% HOF metric, it can also cause UEs to remain connected to a weak-serving BS for too long, leading to outages (RLF) before the next handover occurs (in 4.8\% of the cases, as seen in Fig.~\ref{fig:KPI_compare_HOF_RLF}). 
\item 
In contrast, when optimizing for RLF, the framework accounts for UE conditions both at the time of handover and afterward. This ensures that UEs do not experience outages due to delayed handovers. As a result, optimizing for RLF not only eliminates outages (0\% RLF) but also naturally leads to 0\% HOF, as seen in Fig.~\ref{fig:KPI_compare_HOF_RLF}.
\end{itemize}

\noindent Since minimizing RLF not only reduces outages but also naturally minimizes HOF, in the following, we focus on KPI study \#2, i.e., ping-pongs vs. RLF.

\begin{table}[h]
\centering
\caption{Mobility performance for GUEs at 3\,km/h: 3GPP benchmark configurations (set-1 and set-5), optimizing ping-pongs vs. HOF (`PP-HOF'), and optimizing ping-pongs vs. RLF (`PP-RLF').}
\begin{tabular}{l c c c c}
\hline
 & 3GPP set-1 & 3GPP set-5 & PP-HOF & PP-RLF \\
\hline
Ping-pongs (\%) & \textbf{48.8} & 75.3 & 56.4 & 56.4\\
\hline
HOF (\%) & 6.5 & \textbf{0.0} & \textbf{0.0} & \textbf{0.0}\\
\hline
RLF (\%) & 7.4 & \textbf{0.0} & 4.8 & \textbf{0.0} \\
\hline
\end{tabular}
\label{tab:KPI_performance_compare}
\end{table}

\subsubsection*{All UEs at the same speed}

Fig.~\ref{fig:allcells_vs_percell_pingpongs_RLF_compare} illustrates the mobility performance when all UEs move at the same speed, for different values of such speed (3 km/h, 30 km/h, and 60 km/h). For each value of the speed, the A3-offset and TTT is optimized for ping-pongs and RLF via HD-BO and the objective function weights are set to $w_{\text{PP}} = 9$ and $w_{\text{RLF}} = 1$. The comparison includes a one-threshold optimization approach (where all cells share the same configuration) and a per-cell configuration. 
Compared to the 3GPP set-1 benchmark (optimized for reducing ping-pongs), HD-BO with one-threshold optimization (uniform configuration for all cells) achieves a 20\% reduction in ping-pongs. Per-cell optimization further improves performance, reducing ping-pongs by 35\% for GUEs moving at 3\,km/h. At higher speeds, such as 60\,km/h, the improvement is even more significant, with ping-pong reductions of 57\% and 73\% for one-threshold and per-cell optimization, respectively, compared to the 3GPP benchmark. This is because fast-moving UEs are more likely to traverse multiple cell boundaries within short time intervals, increasing the risk of unnecessary handovers. Adaptive thresholding—especially when done on a per-cell basis—helps mitigate such an effect. These results highlight the advantages of per-cell optimization over one-threshold optimization.

\subsubsection*{Optimizing for the wrong speed}
\begin{figure}
\centering
\includegraphics[width=\figwidth]{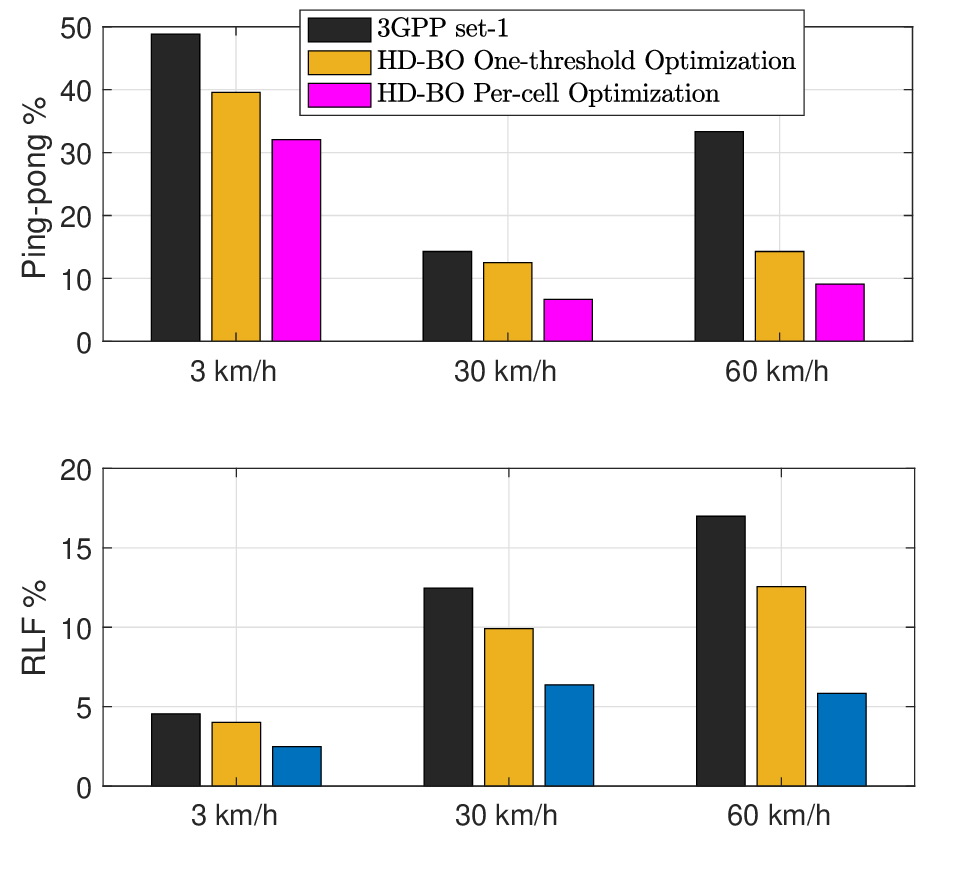}
\caption{GUE ping-pongs and RLF performance at a certain speed when the network is optimized for that speed via HD-BO using one-threshold optimization (uniform configuration for all cells), per-cell configuration, and with the 3GPP set-1 configuration.}
\label{fig:allcells_vs_percell_pingpongs_RLF_compare}
\end{figure}

\begin{figure}
\centering
\includegraphics[width=\figwidth]{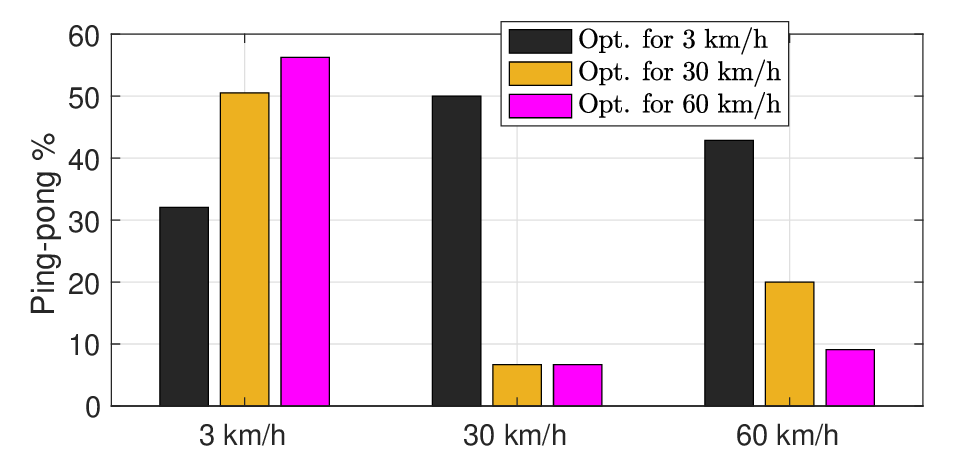}
\caption{GUE ping-pongs at a certain speed when the network is optimized via HD-BO for a specific (not necessarily the same) speed.}
\label{fig:different_speeds_opt_pingpongs_compare}
\end{figure}

Instead, Fig.~\ref{fig:different_speeds_opt_pingpongs_compare} presents the performance for UEs at a certain speed when the handover parameters are optimized for a \emph{different} speed. 
When optimizing for a specific speed, the HD-BO-recommended configuration may degrade the performance of UEs moving at other speeds. For example, optimizing for pedestrian speed (3\,km/h) results in an 8× increase in ping-pongs when the same configuration is applied to higher speeds, such as 30\,km/h. This occurs because TTT plays a critical role at higher speeds, where handovers must be executed in a timely manner before the GUE crosses the cell and moves to a new candidate BS (i.e. the BS providing the highest RSRP). A similar degradation is observed in the reverse scenario, in which optimizing for high-speed mobility (60\,km/h) leads to a 2× increase in ping-pongs when the same configuration is applied to pedestrian mobility (3\,km/h). These findings highlight the necessity to take into account the performance of UEs at all speeds when optimizing the handover parameters, as discussed in the sequel. The substantial gains shown in Fig.~\ref{fig:different_speeds_opt_pingpongs_compare} are a direct consequence of the performance degradations observed when applying speed-mismatched configurations, as evidenced in Fig.~\ref{fig:allcells_vs_percell_pingpongs_RLF_compare}.

\subsubsection*{Optimizing for all speeds}
\begin{figure}
\centering
\includegraphics[width=\figwidth]{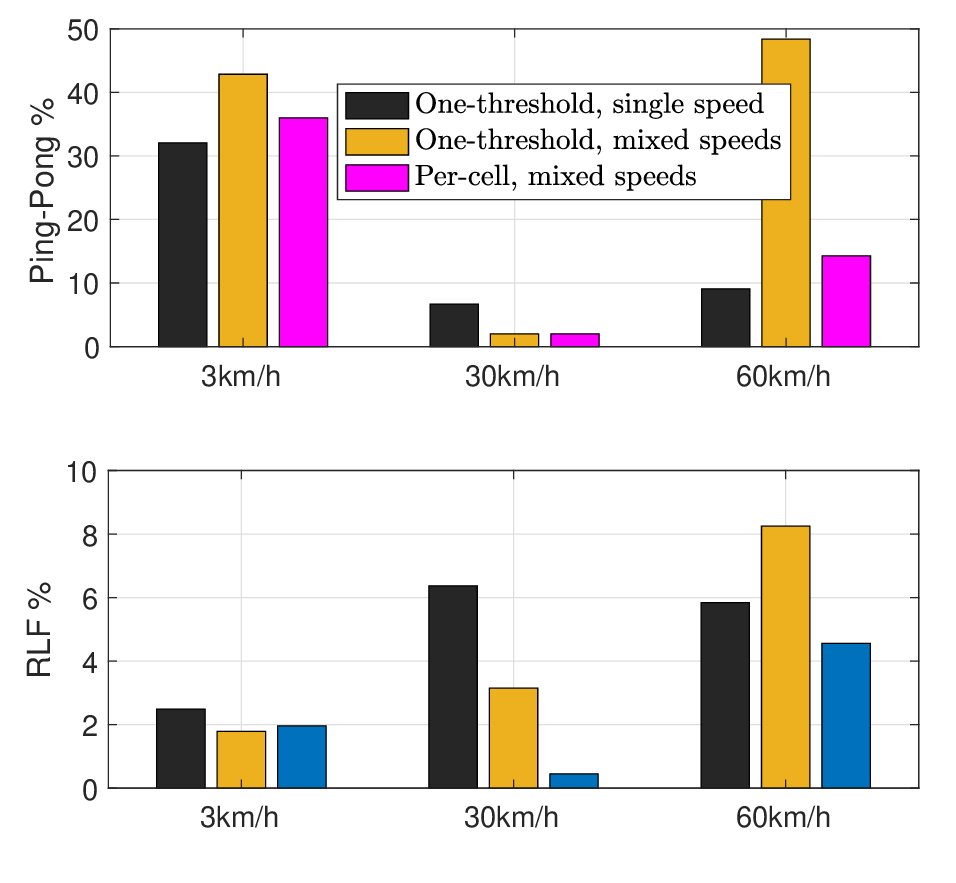}
\caption{Ping-pongs and RLF experienced by GUEs at different speeds when the network is optimized via HD-BO for a mix of all speeds.}
\label{fig:mix_speeds_opt_pingpongs_RLF_compare}
\end{figure}

Fig.~\ref{fig:mix_speeds_opt_pingpongs_RLF_compare} illustrates the performance of UEs at each speed category when the network is optimized for all speeds simultaneously, following per-cell optimization, with $w_{\text{PP}} = 9$ and $w_{\text{RLF}} = 1$. When optimizing for all speeds, GUEs at 60\,km/h experience a slight increase in ping-pongs from 9\% to 14\% (per-cell optimization). However, the impact on 3\,km/h GUEs is significantly reduced, with the increase in ping-pongs constrained from a 2× rise to nearly maintaining the same performance as if the network were optimized solely for that speed category—rising only from 32\% to 36\%. Similar behavior is observed across different speed categories, particularly when HD-BO is applied for per-cell optimization. This performance trade-off arises because, when optimizing for all speeds simultaneously, the configuration cannot perfectly match the requirements of any single speed category, and misalignments similar to those seen in the speed-mismatch scenarios naturally emerge.

\begin{figure}
\centering
\includegraphics[width=\figwidth]{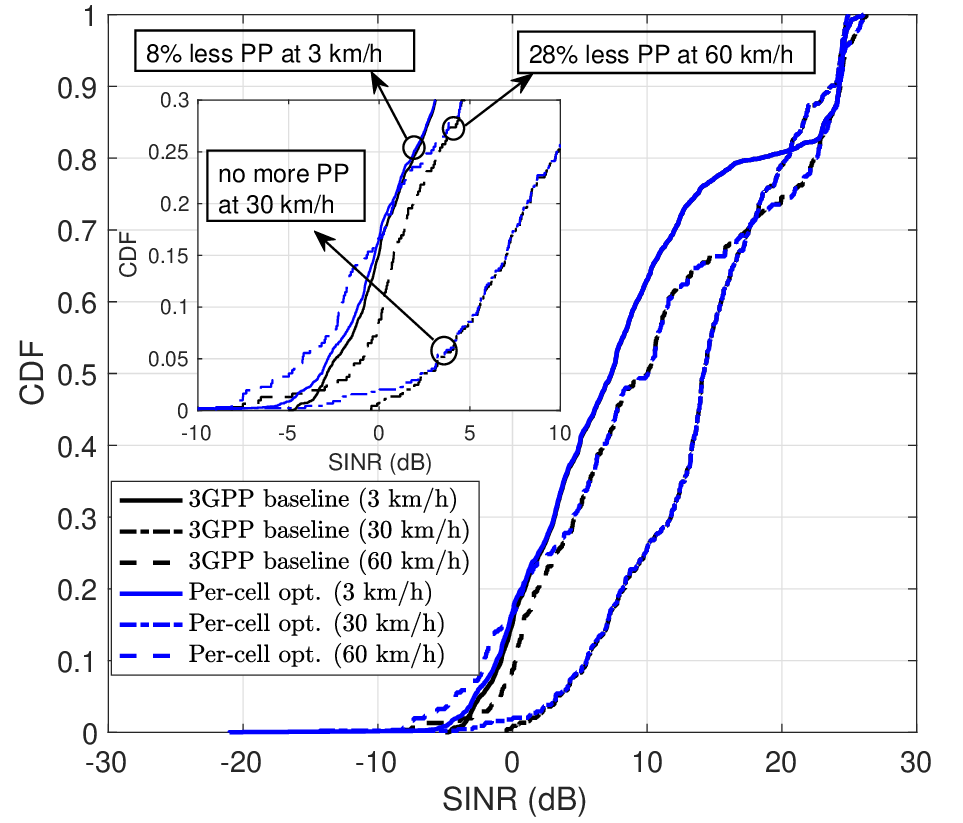}
\caption{SINR achieved by GUEs of different speeds when the network is
optimized for all speeds with $w_{\text{PP}} = 1$ and $w_{\text{RLF}} = 9$, compared to a 3GPP set-5 configuration. Ping-pongs reduction with respect to 3GPP set-5 is indicated in boxed text for each speed.}
\label{fig:SINR_GUE_mixSpeed}
\end{figure}

To further understand the impact of HD-BO on radio link quality, Fig.~\ref{fig:SINR_GUE_mixSpeed} illustrates the CDF of the SINR for UEs at each speed category when the network is optimized for all speeds simultaneously. This figure complements the previous analysis by showing that, despite optimizing handover parameters for reduced RLF and ping-pongs, HD-BO can maintain SINR performance close to the 3GPP set-5 benchmark. The black lines represent the SINR under 3GPP set-5, which serves as an upper bound by enforcing fast handovers to the best candidate BS while ignoring ping-pong constraints. The blue curves show the performance after HD-BO optimization with $w_{\text{PP}} = 1$ and $w_{\text{RLF}} = 9$, prioritizing RLF minimization across a street portfolio with diverse mobility patterns. The similarity in SINR distributions between HD-BO and 3GPP set-5 explains why RLF performance remains nearly identical in both cases: in both strategies, UEs are handed over to strong-signal BSs in a timely manner. Moreover, the reduction in ping-pongs relative to 3GPP set-5 is annotated in boxed text for each speed category—8\% for 3\,km/h, 28\% for 60\,km/h, and 100\% for 30\,km/h—demonstrating that HD-BO successfully mitigates unnecessary handovers without compromising signal quality.

\subsection{Case study \#2: Aerial UE Mobility Management}
\label{sec:case_study_2}

In our second case study, we focus on mobility management for aerial UEs (i.e., UAVs) along predefined 3D aerial corridors at altitudes of 140–160 m.

\subsubsection*{Optimal cellular antenna configurations}

As cellular BSs are traditionally designed to optimize 2D ground-level connectivity, aerial UEs are often limited to receiving signals through the weaker upper antenna sidelobes, resulting in significant signal instability during flight. Additionally, when UAVs fly above buildings, they frequently face interference from line-of-sight (LoS) signals from nearby BSs \cite{GerGarAza2022,GiuNikGer2024}, which degrades their signal-to-interference-plus-noise ratio (SINR) \cite{geraci2018understanding,zeng2020uav}. 
To address this limitation, next-generation mobile networks are expected to provide reliable UAV connectivity by re-engineering existing ground-focused deployments \cite{geraci2022integrating,karimi2023optimizing, KarGerJaf2025}. In our previous study, we proposed a framework for designing a cellular deployment that accommodates both GUEs and UAVs flying along specific streets (i.e., corridors) by optimizing the electrical antenna tilts of each BS. Our findings indicate that, unlike conventional cellular networks where all BSs are downtilted, balancing coverage between ground UEs and UAV corridors necessitates uptilting a subset of BSs \cite{benzaghta2025}. Therefore, for this Case Study \#2 on aerial UE mobility management, before optimizing mobility, we first optimize the electrical antenna tilts of all BSs following the approach in \cite{benzaghta2023designing}. The goal is to
determine the set of antenna tilts that maximize the rates of GUEs and UAVs with equal weights. This ensures an optimized cellular deployment configuration that enhances coverage and capacity for UAVs, serving as a foundation for the subsequent aerial UE mobility optimization.

\subsubsection*{Mobility performance}
Table~\ref{tab:UAVs_comparison} presents the performance of UAVs across all speeds (i.e., 3\,km/h, 30\,km/h, and 60\,km/h), considering an equal weight distribution over the five-street portfolio. The evaluation compares the performance of HD-BO with $w_{\text{PP}} = 9$ and $w_{\text{RLF}} = 1$ to the two 3GPP benchmark configurations, set-1 and set-5. HD-BO outperforms both 3GPP set-1 and set-5 in reducing ping-pongs and RLF. It achieves a 3\% ping-pong rate (compared to 15\% for set-1 and 11\% for set-5) and 0\% RLF (vs. 9\% for set-1). 
Even though the objective function prioritizes ping-pong reduction ($w_{\text{PP}} = 9$, $w_{\text{RLF}} = 1$), HD-BO is still able to completely eliminate RLF ($0\%$), while achieving the best ping-pong performance.

\begin{table}[h]
\centering
\caption{Ping-pong (PP) and RLF performance for UAVs across all speeds with equal weight distribution over the five-street portfolio.}
\begin{tabular}{c c c c c c}
\hline
\multicolumn{2}{c}{3GPP set-1} & \multicolumn{2}{c}{3GPP set-5} & \multicolumn{2}{c}{HD-BO} \\
\hline
PP (\%) & RLF (\%) & PP (\%) & RLF (\%) & PP (\%) & RLF (\%) \\
\hline
15 & 9 & 11 & \textbf{0} & \textbf{3} & \textbf{0} \\
\hline
\end{tabular}
\label{tab:UAVs_comparison}
\end{table}

\subsection{Transfer Learning Experiments}

Since it is desirable for a machine learning model to deliver consistent performance across diverse scenarios \cite{lin2023overview}, we now examine the generalization capabilities of the HD-BO framework across different UE speeds in the context of transfer learning.

\subsubsection*{Scenario source vs. scenario target}
Transfer learning leverages knowledge or data from a previously solved problem (source) to accelerate the solution of a new but related problem (target). This approach is particularly beneficial when generating the initial dataset $\mathcal{D}$ for the BO posterior is costly or time-consuming, such as when real-world measurements are required. 
Let $\mathcal{D}_{\text{sr}}$ and $\mathcal{D}_{\text{tg}}$ represent the initial datasets obtained for the source and target scenarios, respectively. We conduct three evaluations by varying the proportion of the initial dataset $\mathcal{D}$ that originates from the target scenario, as follows:
\begin{itemize}[leftmargin=*]
\item 
100\% ($\mathcal{D} = \mathcal{D}_{\text{tg}}$, prior knowledge based on scenario target).
\item 
50\% (half of $\mathcal{D}$ is drawn from 
$\mathcal{D}_{\text{sr}}$, half is from $\mathcal{D}_{\text{tg}}$).
\item 
0\% ($\mathcal{D} = \mathcal{D}_{\text{sr}}$, prior knowledge based on scenario source).
\end{itemize}

\noindent We apply scenario-specific transfer learning to Case Study \#1, where the objective is to leverage data from a scenario with a certain GUE speed to optimize a new scenario with a different speed.

\subsubsection*{Convergence of transfer learning}

\begin{figure}
\centering
\includegraphics[width=\figwidth]{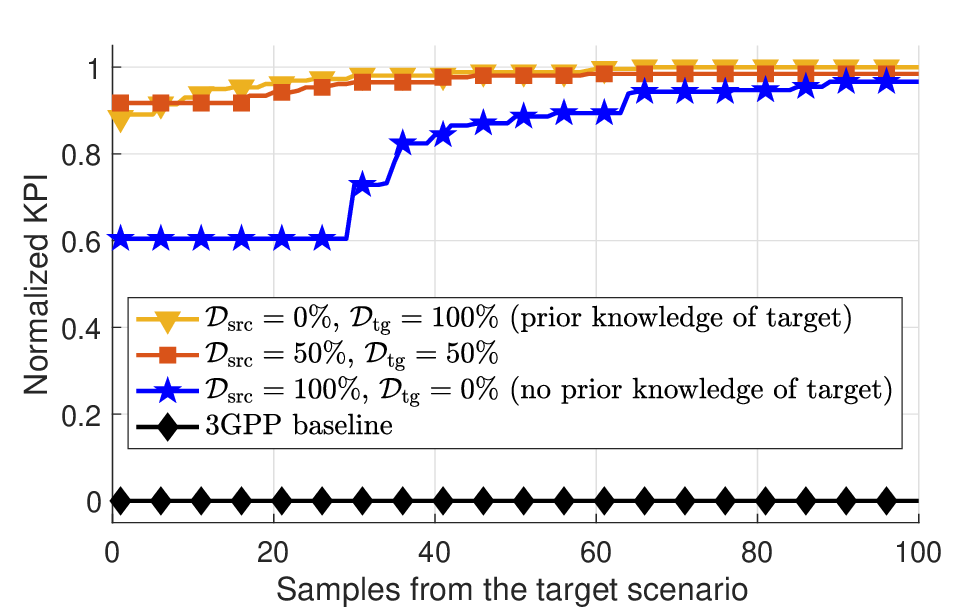}
\caption{Convergence of transfer learning applied on Case Study \#1. Source: All GUEs at 60\,km/h. Target: All GUEs at 30\,km/h. The initial dataset $\mathcal{D}$ contains $N_{\textrm{o}}=60$ observations.}
\label{fig:LT_convg}
\end{figure}

Fig.~\ref{fig:LT_convg} is obtained considering a source scenario based on the previously described Case Study \#1, where GUEs move along a portfolio of five streets at a speed of 60\,km/h. In the target scenario, we modify the GUE speed to 30\,km/h. The figure illustrates the convergence of transfer learning using HD-BO by showing the best observed objective at each iteration $n$. To present a practically relevant measure, we plot the min-max normalized KPI from (3), where 0 corresponds to the 3GPP baseline (worst performance) and 1 represents the best achievable KPI when the HD-BO posterior contains 100\% prior knowledge of the target scenario. The initial dataset $\mathcal{D}$ consists of $N_{\textrm{o}} = 60$ observations, following the recommendation in \cite{eriksson2019scalable}, which suggests building the prior on an initial dataset of size twice the number of optimization parameters. $\mathcal{D}$ is drawn from $\mathcal{D}_{\text{sr}}$ (blue), from $\mathcal{D}_{\text{tg}}$ (green), or half each (red). 
Fig.~\ref{fig:LT_convg} shows that with a 50\%/50\% reliance on 
$\mathcal{D}_{\text{tg}}$/$\mathcal{D}_{\text{sr}}$, convergence occurs in a comparable number of iterations to that observed with 100\% reliance on $\mathcal{D}_{\text{tg}}$ (i.e., without transfer learning). This demonstrates the HD-BO posterior’s ability to generalize after optimizing a related task. Even in the absence of prior knowledge of the target scenario ($\mathcal{D} = \mathcal{D}_{\text{sr}}$), performance declines by only 3\%.
 
\subsubsection*{Performance of successful transfer learning}

\begin{figure}
\centering
\includegraphics[width=\figwidth]{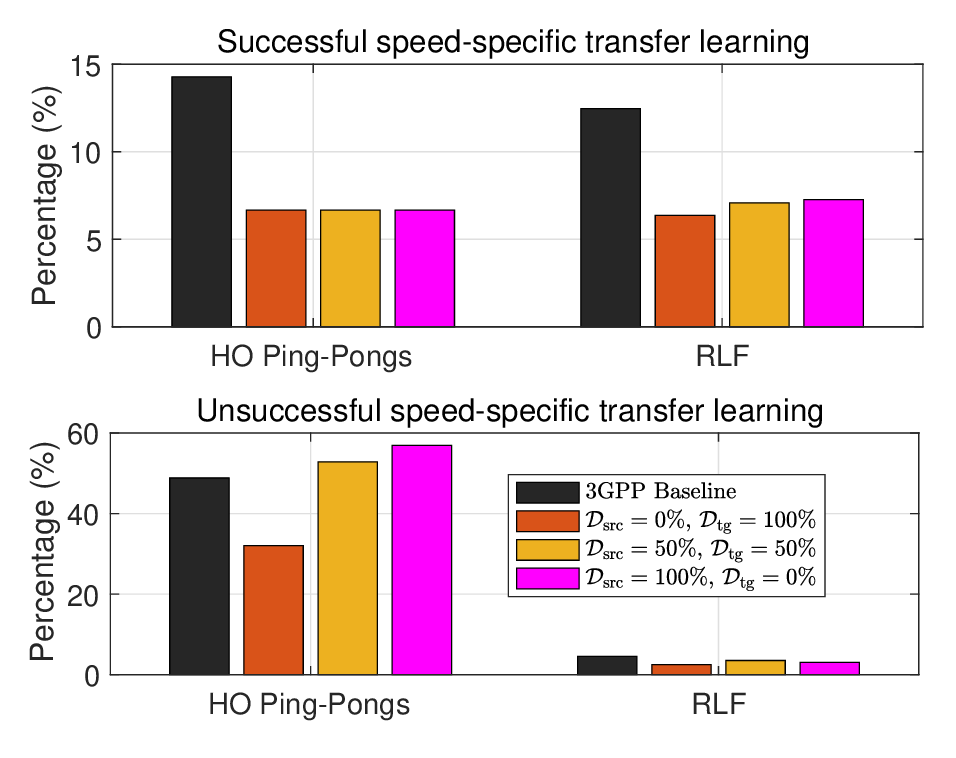}
\caption{Performance of transfer learning applied on Case Study \#1.}
\label{fig:LT_perform}
\end{figure}

Fig.~\ref{fig:LT_perform} compares the achieved ping-pongs and RLF percentages. The results show comparable performance across all three variations of the initial dataset, demonstrating the effectiveness of transfer learning in adapting GUEs to new speed deployments. Specifically, the method successfully generalizes when transferring from a scenario where GUEs move along five streets at 60 km/h to a target scenario at 30 km/h.

\subsubsection*{Example of unsuccessful transfer learning}

However, transfer learning proves ineffective for pedestrian speeds (3 km/h). The figure highlights this limitation, as scenario-specific transfer learning fails in both initial dataset variations (50\%-50\% and 100\% scenario source). In both cases, the achieved ping-pongs and RLF performance do not match the levels obtained when the optimization is conducted with an initial dataset composed entirely of the target scenario (100\% scenario target). This occurs because the HD-BO method establishes trust regions based on presumed solution locations, primarily favoring lower TTT values. Without data specific to pedestrian performance, it lacks awareness of the importance of higher TTT values and consequently defaults to optimizing only for lower ones. As a result, this approach fails to yield improvements for pedestrian-based scenarios.

%% file: 04_RL.tex
\section{Cellular Mobility Management via\\Deep Reinforcement Learning}
\label{sec:RL}
In this section, we present model-free mobility management based on deep reinforcement learning (DRL). Unlike the HD-BO method presented in the previous section, the DRL-based approach does not require predefined parameters for handover decisions. Instead, an agent, deployed at the network side, directly selects the next serving BS for a given UE based on the information obtained from the network state. In our case studies, we compare the KPI performance and convergence of the DRL approach to the one based on HD-BO.

\subsection{Overview of Deep Reinforcement Learning}

RL is a type of machine learning in which an agent learns by interacting with the environment. Using the data collected from these interactions, the agent learns a policy allowing it to select the actions maximizing cumulative rewards \cite{sutton2018}. RL problems are typically modeled as Markov decision processes (MDPs), characterized by states, actions, rewards, state transition dynamics, and a discount factor balancing between immediate and future rewards. At each step, the agent observes a state describing the environment, chooses an action based on its policy (which defines the agent's behavior), and receives an immediate reward. RL algorithms can be classified into two main categories: value-based methods and policy gradient algorithms. Value-based methods, such as Q-learning, rely on value functions which estimates the expected cumulative reward that the agent can achieve, under a specific policy, from a particular state or a state-action pair. These algorithms optimize policies in an implicit fashion by selecting the actions maximizing the estimated value function. Meanwhile, policy gradient methods, including REINFORCE, directly optimize the policy parameters using gradient ascent on the expected reward.

To expand the applicability of RL, deep RL was introduced combining RL with deep neural networks (DNNs) \cite{mnih2015human}. This enables agents to handle more complex environments. Examples of DRL algorithms include Deep Q-Network (DQN), Deep Deterministic Policy Gradient (DDPG), Proximal Policy Optimization (PPO), and Advantage Actor-Critic (A2C). While DQN is designed to deal with discrete actions using DNNs for Q-value function approximation, DDPG is developed to handle continuous actions through an actor-critic architecture. Meanwhile, PPO and A2C use stochastic policies, which not only allow them to make both discrete and continuous actions, but also improve their state space exploration compared to DQN and DDPG. Moreover, as policy-gradient algorithms, PPO and A2C can address the limitations of value-based methods like DQN which suffer from slow convergence and high approximation errors.

\subsection{DRL-based Mobility Management}
In our DRL-based mobility management framework, the state space, action space, and reward function are defined as follows:
\subsubsection*{State space} 
The state $s_t$ of the environment at time step $t$ includes the ID $\Gamma_{t}$ of the street where the UE is located, the ID $\alpha_t$ of the current serving BS, the time of stay at current BS $\text{ToS}_{\alpha_t}$, the ID $\beta_t$ and RSRP $\Xi_t$ of the candidate BS. Additionally, we consider a history $H^n_t$ of $n$ previous serving BSs, defined by the ID $\beta^{i}_t$ and the RSRP value $\Xi^{i}_t$ $i=1 \dots n$, in the state representation. %
%
%
Hence, the state is given by $s_t= \{\Gamma_{t}, \alpha_t, \text{ToS}_{\alpha_t}, \beta_t, \Xi_t, H^n_t\}$.
\subsubsection*{Action space}
The actions of the agent at time step $t$ include a binary action $a_t= \{0,1\}$, representing the decision to stay connected to the current BS or make a handover to the candidate BS. 

\subsubsection*{Reward function}
The reward $r_t$ at time step $t$ is a weighted sum of the ping pong event and the RLF event, given by:
    \begin{equation}
        r_t = -w_{\text{PP}} \cdot \mathds{1}_{\text{PP}_t} - w_{\text{RLF}} \cdot \mathds{1}_{\text{RLF}_t}.
        \label{eqn:RL_Opt_problem} 
    \end{equation}

\subsubsection*{Proximal Policy Optimization}
To solve the formulated RL problem, we adopt PPO \cite{schulman2017proximal}, a DRL algorithm widely used in wireless communication applications, thanks to its robustness, stability, and sample efficiency. PPO optimizes its policy through a clipped objective function, which prevents excessively large policy updates. This offers more stable and reliable training compared to other actor-critic methods. PPO includes two neural networks: the policy network (actor) and the value network (critic). The former is responsible for action selection where the actions are sampled from a probability distribution based on the actor's stochastic policy $\pi_\psi$. Meanwhile, the latter evaluates the actions taken by the policy network through value function estimation. Through interactions between the agent and the environment, batches of trajectories (i.e., sequences of states, actions, rewards, and next states) are collected to update both networks. 

\subsubsection*{Value loss}
The value network parameters $\beta$ are updated by minimizing the value loss, defined as
\begin{equation}
\mathcal{L}_{\text{value}}(\beta)=\mathbb{E}_t\left[\left(V_\beta\left(s_t\right)-G_t\right)^2\right]
\end{equation}
where $V_\beta\left(s_t\right)$ is the estimated state-value function and $G_t$ denotes the observed return. 

\subsubsection*{Policy objective}
Simultaneously, the policy network $\pi_\psi$ is updated by maximizing the policy objective, given by
\begin{equation}
\begin{split}
    \mathcal{L}_{\text{policy}}(\psi)=\mathbb{E}_t[\min(\rho_tA_{\pi_{\psi}}, \operatorname{clip}(\rho_t, 1-\tau, 1+\tau) A_{\pi_{\psi}}] \\ + \epsilon_{\text{exp}} \mathbb{E}_{\pi_\psi}[\log \pi_\psi(a|s_t)]
\end{split}
\end{equation}
where the first term refers to the clipped surrogate objective and the second terms is the entropy loss which encourages exploration. Also, $\rho_t=\frac{\pi_\psi\left(a_t \mid s_t\right)}{\pi_{\psi_{\text {old}}}\left(a_t \mid s_t\right)}$ represents the probability ratio between the new and old policies and $A_{\pi_{\psi}}$ denotes the advantage estimate, which measures the quality of an action given a state based on the critic's estimated value function. In addition, $\epsilon_{\text{exp}}$ and $\tau$ are the entropy coefficient and the clipping hyperparameter, respectively.

\subsubsection*{Training process}
The training process is repeated over multiple episodes, allowing the critic to improve its value function estimation and the actor to enhance its action selection based on the critic's feedback. 
We fine-tune the different hyperparameters of the PPO algorithm (including both the actor and the critic networks architectures) through extensive experimentation. The policy network is designed using three layers with 256, 128, and 256 neurons, respectively, while the value network includes three layers with 64 neurons each. To optimize the losses, an Adam optimizer is used with a learning rate of 0.0001. Additionally, we select a discount factor $\gamma=0.95$, an entropy coefficient $\epsilon_{\text{exp}}=0.002$, and a clipping parameter $\tau=0.2$.

\subsection{Performance and Convergence Assessment}

We now compare the performance of the PPO agent to the HD-BO approach and to the 3GPP baseline configurations (set-1 and set-5) for Case Study \#1, i.e., GUE mobility management. The PPO agent's objective is to maximize the reward function defined in (\ref{eqn:RL_Opt_problem}). Unlike traditional approaches, the RL-PPO agent selects the next serving BS without relying on predefined network parameters such as the A3-offset and TTT. 

\begin{figure}
\centering
\includegraphics[width=\figwidth]{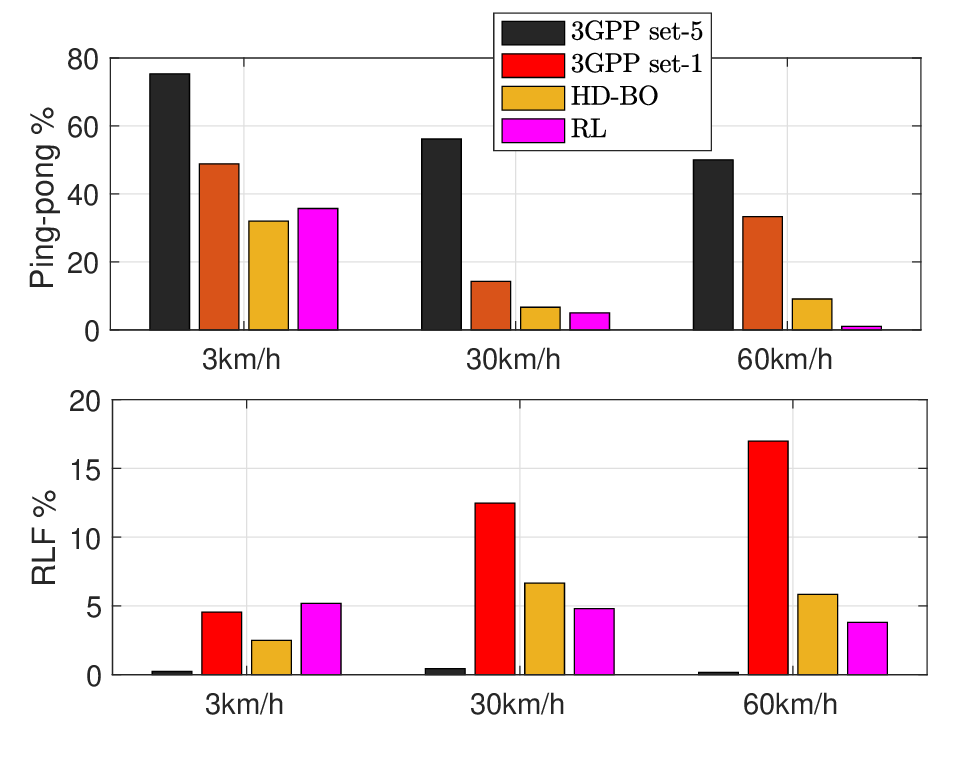}
\caption{Ping-pongs and RLF performance for GUEs at different speeds when the network is optimized via RL-PPO, HD-BO, and the 3GPP baseline configurations.}
\label{fig:RL_pingpongs_RLF_compare}
\end{figure}

Fig.~\ref{fig:RL_pingpongs_RLF_compare} illustrates the mobility performance at different speeds (3 km/h, 30 km/h, and 60 km/h) in terms of ping-pongs and RLF. The objective function weights are set to $w_{\text{PP}} = 9$ and $w_{\text{RLF}} = 1$. The following key observations can be drawn.
\subsubsection*{RL-PPO vs. 3GPP baselines}
The RL-PPO framework significantly outperforms both 3GPP set-1 and set-5 configurations in terms of reducing ping-pongs. For example, at 30 km/h, RL-PPO achieves a ping-pong rate of 5\%, compared to 56.16\% for set-5 and 14.2\% for set-1. Additionally, RL-PPO maintains a lower RLF rate across all speeds, despite the objective function prioritizing ping-pong reduction by assigning it a higher weight. This demonstrates RL-PPO's ability of providing robust mobility management.%
\footnote{3GPP set-5 can be regarded as a practical upper bound in terms of RLF performance, as it performs fast handovers without considering the ping-pong effect, focusing solely on minimizing RLF. Therefore, the primary comparison should be made with 3GPP set-1, which is specifically designed to reduce ping-pongs while also accounting for RLF.}

\subsubsection*{RL-PPO vs. HD-BO}
The performance achieved through RL-PPO is comparable to HD-BO in both ping-pong reduction and RLF minimization. For instance, at 3\,km/h, RL-PPO maintains a ping-pong rate of 35\%, similar to HD-BO’s 32\%. At 30\,km/h RL-PPO performs slightly better achieving a ping-pong rate of 5\%, compared with HD-BO’s 7\%. At 60\,km/h, RL-PPO achieves a ping-pong rate of nearly 0\%. Performance-wise, this confirms RL-PPO as a viable alternative to parameter-based mobility management, without predefined A3-offset and TTT thresholds.

\subsubsection*{Sample efficiency}
Table~\ref{tab:HD_BO_RL_comparison} compares the convergence behavior of HD-BO and the RL-PPO framework in terms of the total number of iterations required. This comparison is conducted for both objective function weight configurations: \{$w_{\text{PP}} = 9$, $w_{\text{RLF}} = 1$\} and \{$w_{\text{PP}} = 1$, $w_{\text{RLF}} = 9$\}. While RL-PPO achieves comparable performance to HD-BO, a key drawback is its significantly higher sample complexity. Each iteration in both methods involves running a simulation or taking a measurement, making sample efficiency critical. RL-PPO requires 10 to 250 times more iterations before convergence compared to HD-BO, depending on the KPI weight configuration and GUE speed. For instance, at 30 km/h with $w_{\text{PP}} = 9, w_{\text{RLF}} = 1$, RL-PPO requires 14,000 iterations—each corresponding to a costly simulation—whereas HD-BO converges in just 60 iterations.

Although RL-PPO is well-suited for digital twin environments, where large-scale simulations can efficiently generate training data, in real-world measurement-based scenarios, where data collection is mission-critical, costly, or labor-intensive, its high sample complexity makes RL-PPO less viable compared to HD-BO's data-efficient optimization approach.



\begin{table}[h]
\centering
\caption{Convergence comparison based on number of iterations at different speeds with varying KPI weight parameters.}
\resizebox{\columnwidth}{!}{%
\begin{tabular}{l c c c c c c}
\hline
 & \multicolumn{2}{c}{3 km/h} & \multicolumn{2}{c}{30 km/h} & \multicolumn{2}{c}{60 km/h} \\
\hline
 & HD-BO & RL-PPO & HD-BO & RL-PPO & HD-BO & RL-PPO \\
\hline
$w_{\text{PP}} = 9$, $w_{\text{RLF}} = 1$ & \textbf{125} & 3200 & \textbf{60} & 14000 & \textbf{30} & 4300 \\
$w_{\text{PP}} = 1$, $w_{\text{RLF}} = 9$ & \textbf{140} & 320 & \textbf{75} & 300 & \textbf{45} & 270 \\
\hline\end{tabular}%
}
\label{tab:HD_BO_RL_comparison}
\end{table}

\subsection{Transfer Learning Experiments}

We now evaluate the generalization capability of the RL-PPO-based mobility management framework in adapting to aerial UEs at a altitude of 150\,m, using an agent trained exclusively on GUEs at 1.5\,m. The study focuses on a mobility scenario where both categories of UEs move at a speed of 30\,km/h. As previously discussed, RL-PPO-based mobility management requires large-scale simulations to generate extensive training datasets. The objective of transfer learning in this context is to minimize the need for collecting new data when UE altitude changes (e.g., when an aerial highway is relocated due to regulations \cite{bernabe2024massive}). 

Table~\ref{tab:TR_RL_UAVs} presents the PP and RLF performance for $w_{\text{PP}} = 9$ and $w_{\text{RLF}} = 1$. The results highlight the effectiveness of transfer learning in reducing training overhead while maintaining optimal mobility performance. With transfer learning, both PP and RLF rates remain at 0\%, while significantly reducing the number of iterations required for training. 
Without transfer learning, the RL-PPO agent requires 6,200 iterations to converge. Transfer learning cuts this down to 2,400 iterations---a 2.5× reduction in training effort---successfully generalizing to aerial UEs without compromising handover efficiency. 

\begin{table}[h]
\centering
\caption{PP and RLF performance for UAVs at 150\,m and 30\,km/h, w/ and w/o RL-PPO transfer learning, for $w_{\text{PP}} = 9$ and $w_{\text{RLF}} = 1$.}
\begin{tabular}{l c c c c}
\hline
 & Set-1 & Set-5 & w/o Transfer Learning & w/ Transfer Learning \\
\hline
PP (\%) & 0.0 & 20.0 & 0.0 & 0.0\\
\hline
RLF (\%) & 5.18 & 0.0 & 0.0 & 0.0\\
\hline
Iterations & - & - & 6200 & 2400 \\
\hline
\end{tabular}
\label{tab:TR_RL_UAVs}
\end{table}

%% file: 05_Conclusion.tex
\section{Conclusion}
\label{sec:conclusion}
This paper explored data-driven approaches for mobility management in cellular networks using two methodologies, HD-BO and DRL. Both approaches aim to optimize HO performance, targeting trade-offs between ping-pongs and RLF. The HD-BO method enables scalable, sample-efficient optimization over large cellular deployments by constructing local surrogate models and leveraging trust-region strategies. It demonstrates strong performance across diverse UE mobility profiles, including both GUEs and UAVs, without requiring extensive training data. In contrast, DRL provides a model-free solution that bypasses the need for predefined HO thresholds, directly learning optimal mobility policies from interaction with the environment. While DRL offers flexibility, it requires significantly more training iterations, which can limit its feasibility in real-world deployments where data collection is costly or constrained. To address this, we apply transfer learning to both approaches, demonstrating its ability to accelerate convergence and improve generalization across UE speeds and altitudes. These results suggest that transfer learning is particularly effective in reducing training demands, especially for DRL, and highlight the broader potential of adaptive, data-driven mobility management in dense and heterogeneous network scenarios.


Several areas remain open for further exploration, including the following ones:
\begin{itemize}
    \item \emph{Beam-based mobility management:} Future 6G deployments may operate in the FR3 spectrum and rely on highly directional beams. Investigating beam-based HO strategies that integrate beam selection and mobility optimization is crucial for ensuring seamless connectivity.
    \item \emph{Multi-RAT handovers:} In integrated terrestrial and non-terrestrial networks (NTN), UEs may switch between cellular and satellite network segments \cite{benzaghta2022uav}. Extending our framework to multi-RAT handovers would enable load balancing and mobility robustness across a heterogeneous infrastructure. 
    \item \emph{Multi-agent RL:} Extending the DRL-based HO management to multi-agent systems could improve performance in large-scale networks with load balancing-aware mobility management. Specifically, in such scenario the state and action spaces would be extremely large for one agent to handle. Thus, multiple RL agents could be deployed where each agent has partial knowledge of the network. Then, using a distributed learning approach, HO management could be optimized with reduced overhead.   
    
\end{itemize}